\documentclass[epj]{svjour}
\usepackage{amssymb}
\usepackage{epsfig}
\begin{document}
%
\title{Longitudinally Polarized Photoproduction of Heavy Flavors at Next-to-Leading Order of QCD}
\author{Johann Riedl\inst{1} \and Marco Stratmann\inst{2} \and
Andreas Sch\"afer\inst{1}}
\institute{Institut f\"ur Theoretische Physik, Universit\"at Regensburg, 93040 Regensburg, Germany
\and
Physics Department, Brookhaven National Laboratory, Upton, NY 11973, USA}
\date{}
\abstract{
We present a phenomenological study of charm quark photoproduction 
in longitudinally polarized lepton-hadron collisions at 
next-to-leading order accuracy of QCD.
Our results are based on a recently developed,
flexible parton-level Monte Carlo program for spin-dependent
heavy flavor hadroproduction, which we
extend to deal also with both direct and resolved photon contributions.
The subsequent hadronization into charmed mesons is modeled
in our calculations, which allows us to compare with data 
on double-spin asymmetries for $D^0$ meson production 
taken by the COMPASS collaboration.
In general, next-to-leading order QCD corrections are found to be 
very significant and do not cancel in spin asymmetries.
We elucidate the role of the individual hard scattering subprocesses
and determine the range of parton momentum fractions 
predominantly probed for charm production at COMPASS.
Theoretical uncertainties are estimated by varying renormalization and
factorization scales and parameters controlling the hadronization
of the charm quarks.
}
\maketitle
%
\section{Motivation and Introduction}
%
The quest to understand the partonic structure of the nucleon spin 
remains to be one of the key research areas in Hadronic Physics even
after more than 25 years of strenuous experimental and theoretical efforts.
In particular, the polarized gluon density, defined as
\begin{equation}
\label{eq:gluon}
\Delta g(x,\mu) \equiv g_+(x,\mu) - g_-(x,\mu)\;,
\end{equation}
is still one of the most elusive quantities associated with
the non-perturbative partonic structure of hadrons.
Here, $g_+$ ($g_-$) denotes the probability of finding
a gluon at a scale $\mu$ with light-cone momentum fraction $x$ 
and helicity $+$ ($-$) in a proton with helicity $+$.
The total, $x$ integrated gluon polarization,
\begin{equation}
\label{eq:firstmom}
\Delta g(\mu) \equiv \int_0^1 \Delta g(x,\mu)\; dx\;,
\end{equation}
and a similar contribution from the sum of all quarks and antiquarks,
enters the helicity sum rule of the nucleon along with 
the orbital angular momenta of quarks and gluons \cite{ref:int-oam}.
The challenge is to precisely map $\Delta g(x,\mu)$
in a wide range of $x$ in order to minimize extrapolation
uncertainties in the first moment (\ref{eq:firstmom}).  

Currently, the best constraints on $\Delta g(x,\mu)$ are derived from global
QCD analyses \cite{deFlorian:2008mr,Blumlein:2010rn} which
treat all available experimental probes simultaneously and
consistently at a given order in the strong coupling 
$\alpha_s$ in perturbative QCD.
The availability of next-to-leading order (NLO) QCD corrections
is essential for any meaningful, quantitative analysis of parton densities.
In particular, single-inclusive pion \cite{Jager:2002xm} and jet production \cite{Jager:2004jh}, 
measured in spin-de\-pen\-dent proton-proton collisions at BNL-RHIC \cite{ref:rhic}, 
have started to put significant limits on the amount of 
gluon polarization in the nucleon \cite{deFlorian:2008mr}.
New, preliminary single and di-jet data from the STAR collaboration \cite{ref:starnew} 
show for the first time tantalizing hints for a non-zero $\Delta g(x,\mu)$ \cite{ref:dssvplus,ref:rhicwp}. 
Due to the given kinematics, the current probes mainly constrain
$\Delta g$ in the medium-to-large $x$ region, $0.05 \lesssim x \lesssim 0.2$,
which is not sufficient to reliably determine its integral (\ref{eq:firstmom}).
A very significant contribution of up to one unit of $\hbar$,
i.e., twice the proton spin, can still come from the unexplored 
small $x$ region \cite{deFlorian:2008mr}.
Narrowing down the uncertainties on $\Delta g(x,\mu)$
and, at the same time, extending the range in $x$ 
continues to be the main objective of experimental efforts in the years to come.
Eventually, only a future high-energy polarized electron-proton collider, 
such as the EIC project \cite{Boer:2011fh}, will finally be able to quantitatively 
address all the remaining open questions related to the 
helicity structure in the small $x$ region \cite{Aschenauer:2012ve}. 
 
\begin{figure}[th]
\vspace*{-0.75cm}
\includegraphics[width=0.52\textwidth]{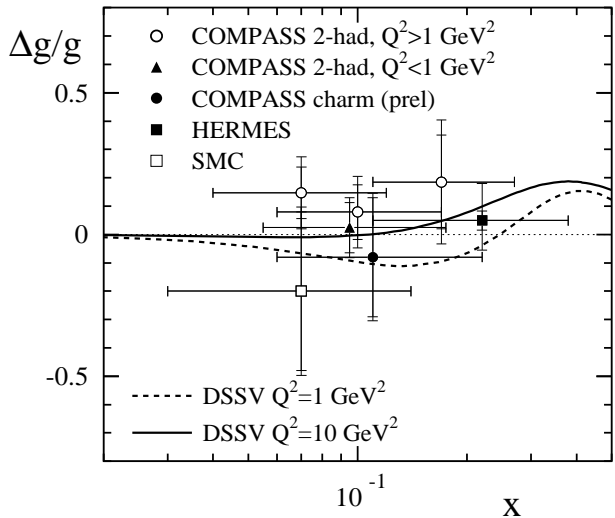}
\vspace*{-0.75cm}
\caption{\label{fig:deltag}
Approximate LO extractions of $\Delta g/g$ from polarized lepton-nucleon scattering
experiments \cite{ref:hermes,ref:smc,ref:compass-2had,ref:compass-charm} 
compared to results of the DSSV global analysis 
of helicity PDFs \cite{deFlorian:2008mr}
based on RHIC $pp$ data for two different scales $Q$.}
\end{figure}
World-data on polarized inclusive and semi-inclusive deep-inelastic scattering 
are pivotal in constraining the helicity quark and antiquark densities 
\cite{deFlorian:2008mr,Blumlein:2010rn,ref:dssvplus},
but due to lack of sufficient kinematic coverage of present fixed-target
experiments, information on the gluon density is notoriously difficult to obtain 
from QCD scaling violations.
Viable probes of $\Delta g(x,\mu)$ at fixed-target energies
comprise of one- and two-hadron and open charm production
and have been exploited by several experimental collaborations
\cite{ref:hermes,ref:smc,ref:compass-2had,ref:compass-charm}.
The proper theoretical description of these processes depends on the
virtuality of the probing photon and is in general more involved than
corresponding calculations for hadron-hadron scattering.
In case of photoproduction, where a quasi-real photon is exchanged,
one has to include also ``resolved'' contributions, where the photon 
fluctuates into a vector meson of the same quantum numbers
before the hard scattering with partons in the proton takes place.
If the virtuality $Q$ of the photon is of ${\cal{O}}(1\,\mathrm{GeV})$
or higher, resolved processes are sufficiently suppressed but the
additional momentum scale $Q$ greatly complicates the calculations
of phase-space and loop integrals.

Not surprisingly, only very few calculations at NLO of QCD are available in 
case of polarized beams and targets \cite{Bojak:1998bd,Merebashvili:2000ya,Jager:2003vy}, 
none of which for large virtualities $Q$. 
As a consequence, the available data sets on hadron and open charm
production \cite{ref:hermes,ref:smc,ref:compass-2had,ref:compass-charm}
have not been included in global QCD analyses of helicity parton densities so far.
Experiments have analyzed their data only in terms of the gluon polarization,
$\Delta g(x,\mu)/g(x,\mu)$, under certain simplifying assumptions and based on leading order (LO)
matrix elements. Nonetheless, the results of these exercises, illustrated in Fig.~\ref{fig:deltag},
are in fairly good agreement with a NLO extraction of $\Delta g(x,\mu)$ including 
the RHIC $pp$ data \cite{deFlorian:2008mr}.
While determinations of $\Delta g$ at fixed-target energies are more involved and
provide less of a constraint than collider data, they are crucial for further testing and establishing the 
assumed universality of helicity-dependent parton densities and hence for 
our understanding of the spin structure of the nucleon and QCD in general.
On the one hand, lepton-nucleon scattering experiments are sensitive to 
different partonic hard scattering processes than jet or hadron production at RHIC and, 
on the other hand, the relevant momentum fractions $x$ fall within the range already
probed by $pp$ data.

In this paper, we present a comprehensive phenomenological analysis of 
longitudinally polarized photoproduction of heavy flavors at NLO of QCD.
To this end, we extend our recently developed flexible parton-level Monte Carlo program
for spin-dependent heavy flavor hadroproduction \cite{Riedl:2009ye} by including 
all relevant subprocesses for direct \cite{Bojak:1998bd,Merebashvili:2000ya}
and resolved \cite{Riedl:2009ye,Bojak:2001fx} photon contributions.
To facilitate the comparison to data, the hadronization of the produced charm
quarks into charmed mesons and, as an additional option, their semi-leptonic decays can be
modeled in our calculations based on the phenomenological functions 
used in Ref.~\cite{ref:cacciari}.
The presented results will allow one to consistently include available and upcoming data
on spin-dependent open charm photoproduction into future global QCD analyses
of helicity parton densities at NLO accuracy.

As a first application, we examine in detail all aspects of open charm production
in the kinematic regime accessible with the COMPASS experiment \cite{ref:compass-charm}.
We will demonstrate that NLO corrections are very significant and do not cancel in
double-spin asymmetries used to extract information on the polarized gluon density
$\Delta g(x,\mu)$.
We elucidate the relevance of the individual direct 
and resolved partonic subprocesses and estimate theoretical uncertainties
by varying renormalization and factorization scales and parameters controlling 
the hadronization of the produced charm quarks into experimentally observed $D$ mesons.
Finally, we shall illustrate the sensitivity of the 
existing data on the double-spin asymmetry for charm quark photoproduction 
\cite{ref:compass-charm,ref:kurek,ref:franco-phd,ref:franco}
to $\Delta g(x,\mu)$ and compare our estimates of the relevant range of momentum fractions $x$ 
with those obtained by the COMPASS collaboration. 

We note that our flexible Monte Carlo code is capable of computing any infrared safe
heavy flavor photoproduction cross section at ${\cal{O}}(\alpha\alpha_s^2)$
in longitudinally polarized lepton-nucleon collisions, including
correlations of the produced heavy quark pair. Our results complement and significantly extend 
previously existing spin-dependent NLO calculations of single-inclusive heavy quark yields
based on largely analytical methods \cite{Bojak:1998bd,Merebashvili:2000ya}, 
where the resolved contribution was neglected, any information on the partonic recoil system was lost, 
and most experimental cuts could not be implemented.
The results presented here will be also useful for future studies of 
spin-dependent charm photoproduction at an EIC \cite{Boer:2011fh}.

The outline of the paper is as follows: in Sec.\ II we briefly 
review some of the technical aspects of setting up a parton-level Monte Carlo program
for heavy flavor photoproduction in polarized lepton-nucleon collisions at NLO accuracy.
In Sec.\ III we present a detailed phenomenological study of open charm quark production
at COMPASS, including the relevance of NLO corrections, the role of
the different hard scattering subprocesses, an estimate of the relevant momentum fractions $x$,
and an assessment of theoretical uncertainties.
We summarize our results in Sec.\ IV.

\section{Technical Framework\label{sec:tech}}
%
We consider heavy quark photoproduction in longitudinally polarized 
lepton-nucleon collisions at 
NLO accuracy of QCD by consistently including for the first time 
both direct and resolved photon contributions.
In order to compute arbitrary infrared-safe observables
within flexible experimental acceptance cuts and to account for
the hadroni\-zation of the produced heavy quark pair, 
all phase-space integrations are 
performed numerically with appropriate Monte Carlo techniques.
To this end, we follow closely the subtraction method devised and used in
Refs.~\cite{Mangano:1991jk,Frixione:1993dg}. In the following, we
will only briefly outline the technical details relevant 
for the case of polarized photoproduction.

Assuming, as usual, factorization, the inclusive cross section 
for producing a heavy quark $Q$ in 
spin-dependent lepton-pro\-ton collisions at a 
center-of-mass system (c.m.s.) energy $\sqrt{S}$ 
can be written as a convolution, 
\begin{eqnarray}
\label{eq:xsec-def}
\nonumber
d\Delta\sigma^Q &\equiv& \frac{1}{2} \left[
d\sigma^Q_{++} - d\sigma^Q_{+-}\right]
\\
\nonumber
&=& \sum_{a,b}\int dx_a dx_b \Delta f_a^l(x_a,\mu_f) \Delta f_b^p(x_b,\mu_f)\,{\cal{S}}\\
&\times& d\Delta\hat{\sigma}_{ab}(x_a,x_b,S,m_Q,k_1,k_2,\mu_f,\mu_r)\,,
\label{eq:xsec-fact}
\end{eqnarray}
where the subscripts $\pm$ in (\ref{eq:xsec-def}) 
label the helicity states of the lepton and proton.
In analogy to Eq.~(\ref{eq:gluon}),
the $\Delta f^{l,p}_{a,b}$ denote the spin-dependent 
parton distribution functions (PDFs) of flavor $a,b$  
in a lepton $l$ or proton $p$.
The sum in (\ref{eq:xsec-fact}) is over all contributing partonic processes
$ab\rightarrow Q\bar{Q}c$, 
including those with a direct photon, i.e.,
$a=\gamma$, with $d\Delta\hat{\sigma}_{ab}$ the
associated polarized hard scattering cross sections.
The required spin-dependent matrix elements squared $\Delta |M_{ab}|^2$
at NLO  accuracy in $d=4-2\varepsilon$ dimensional regularization 
for the direct and resolved photon processes can be taken from
\cite{Bojak:1998bd} and \cite{Bojak:2001fx}, respectively.
Corresponding unpolarized results can be found in
Refs.~\cite{Mangano:1991jk,Frixione:1993dg,ref:nlo-unpol}.
At ${\cal{O}}(\alpha\alpha_s^2)$, parton $c$ 
can be either a gluon or a light (anti-)quark producing an additional jet
or hadron which is usually not observed in experiment.
$k_{1,2}$ denote the four-momenta of the heavy quark $Q$ and antiquark $\bar{Q}$ 
with mass $m_Q$, i.e., $k_{1,2}^2=m_Q^2$. 
$\mu_r$ and $\mu_f$ are the renormalization and factorization scale,
respectively, which are typically chosen as a combination of the
hard scales characterizing the process.

The ``measurement function'' ${\cal{S}}$ in Eq.~(\ref{eq:xsec-fact})
defines the observable, for instance, through a set of
step functions implementing the experimental cuts imposed
on the final-state particles and selecting a certain bin in
a kinematical distribution of interest.
In case of the COMPASS experiment, 
single-inclusive spectra of $D$ mesons, 
differential in transverse momentum $p_T^D$,
are measured \cite{ref:compass-charm,ref:kurek,ref:franco-phd,ref:franco}. 
Thus, the cross section (\ref{eq:xsec-fact}) at the heavy quark-level is
not yet sufficient for comparing theory with experimental results
and needs to be convoluted with an additional
phenomenological function $D^{Q\rightarrow H_Q}$ 
modeling the hadronization of the charm quark:
\begin{equation}
\label{eq:conv}
d\Delta \sigma^{H_Q} = d\Delta \sigma^Q \otimes D^{Q\to H_Q}\;.
\end{equation}
We will specify our choice for $D^{Q\rightarrow H_Q}$,
along with all other non-perturbative inputs, in Sec.~\ref{sec:pheno}.
Even though not relevant for the phenomenological
applications considered in this paper, our
flexible Monte Carlo program can also
account for the semi-leptonic decays of $D$ mesons
if needed.

For the integration of the fully exclusive partonic cross sections
$d\Delta\hat{\sigma}_{ab}$ in Eq.~(\ref{eq:xsec-fact}), 
we generalize the framework used for unpolarized heavy quark
production in \cite{Mangano:1991jk,Frixione:1993dg}
to deal with singular regions of phase-space.
The gist of the method is to add and subtract appropriate 
spin-dependent counter terms to $d\Delta\hat{\sigma}_{ab}$
which can be integrated analytically with respect to
momenta of unresolved partons.
For all resolved photon processes, $a\neq \gamma$, 
we can adopt the expressions obtained in our recent calculation of polarized heavy
quark hadroproduction at ${\cal{O}}(\alpha_s^3)$ \cite{Riedl:2009ye},
since these processes share the same hard scattering matrix elements.
Hence, we only need to consider processes with a direct photon,
$d\Delta\hat{\sigma}_{\gamma b}$, in the following.

In general, for a numerically efficient implementation of
the subtraction method it is convenient \cite{Mangano:1991jk,Frixione:1993dg,Riedl:2009ye}
to express the three-body phase-space in terms of
variables where soft and collinear singularities are particularly transparent.
This is achieved by choosing $x=(k_1+k_2)^2/s$, the invariant mass of the
$Q\bar{Q}$ pair scaled by the available partonic c.m.s.\ energy squared, 
i.e., $4m_Q^2/s \le x \le 1$. In addition, one uses $-1\le y \le 1$, 
the cosine of the angle between the $z$-direction,
aligned with the spatial direction of parton $a$, and $\vec{k}_3$, 
the momentum of parton $c$, in the c.m.s.~of the incoming partons.
Soft and collinear regions of phase-space are then simply associated with
$x=1$ and $y=\pm 1$, respectively, and the hard scattering matrix element 
for the process $\gamma b\to Q\bar Q c$ can be written as
\begin{equation}
\label{eq:m-sing}
\Delta|M_{\gamma b}|^2=\frac{\Delta f_{\gamma b}(s,m_Q,x,y,\theta_1,\theta_2)}{s^2(1-x)^2(1-y^2)}\,,
\end{equation}  
where $\Delta f_{\gamma b}$ is regular for $x=1$ and $y=\pm1$.
The angles $\theta_{1,2}$ are used to parametrize the spatial orientation of
$k_{1,2}$ with respect to the plane span by the other three momenta in
the c.m.s.\ of the $Q\bar{Q}$ pair;
for further details, see 
\cite{Mangano:1991jk,Frixione:1993dg,Riedl:2009ye}.
We note that the genuine NLO subprocess $\gamma q (\bar{q}) \to Q\bar{Q} q (\bar{q})$
can have only collinear singularities at ${\cal{O}}(\alpha\alpha_s^2)$.

The partonic subprocesses contributing to the direct photon
cross section at ${\cal{O}}(\alpha \alpha_s^2)$,
$\gamma g\to Q\bar{Q}g$ and $\gamma q\, (\bar{q}) \to Q\bar{Q} q\, (\bar{q})$,
can be decomposed as \cite{Frixione:1993dg}
\begin{eqnarray}
\label{eq:xsec-decomp}
\nonumber
d\Delta\hat{\sigma}_{\gamma b} &=&
d\Delta\hat\sigma^{(b)}_{\gamma b}
+d\Delta\hat{\sigma}^{(c+)}_{\gamma b}+d\Delta\hat{\sigma}^{(c-)}_{\gamma b}\\
&+&d\Delta\hat{\sigma}^{(s)}_{\gamma b}+d\Delta\hat{\sigma}^{(v)}_{\gamma b}+d\Delta\hat{\sigma}^{(f)}_{\gamma b}.
\end{eqnarray}
Here, $d\Delta\hat\sigma^{(b)}_{\gamma b}$ and $d\Delta\hat{\sigma}^{(v)}_{\gamma b}$
denote the ${\cal{O}}(\alpha \alpha_s)$ Born contribution and the ${\cal{O}}(\alpha \alpha_s^2)$ 
one-loop corrections to the $\gamma g\to Q\bar{Q}$ process, respectively.
Analytic expressions for the virtual contributions in $d$ dimensions, with ultraviolet divergences 
being subtracted at the renormalization scale $\mu_r$, can be found in Ref.~\cite{Bojak:1998bd}.

In Eq.~(\ref{eq:xsec-decomp}), 
$d\Delta\hat{\sigma}^{({s})}_{\gamma b}$ 
denotes the soft gluon emission part
of the $\gamma g$ scattering cross section, which can be obtained
from the full $d$-dimensional matrix elements squared
in the limit $x\rightarrow 1$ where the phase-space 
integrations can be performed analytically \cite{Frixione:1993dg}.
We obtain the same result for the squared three-body amplitude in the
soft limit as given in Eq.~(A.14) of Ref.~\cite{Frixione:1993dg}
but with the Born contribution  
for $\gamma g \to Q\bar{Q}$ being replaced by its spin-dependent,
color-averaged counterpart, which reads in 
$d=4-2\varepsilon$ dimensions:
\begin{equation}
\label{eq:mgglo}
\Delta|M_{\gamma g}|^2 = \frac{1}{2s}  (4\pi\alpha_s) e^2 e_Q^2 
\Delta B_{\gamma g}
\end{equation}
where
\begin{equation}
\label{eq:bqed}
\Delta B_{\gamma g} = \left(\frac{t_1}{u_1}+\frac{u_1}{t_1}\right)
\left(\frac{2 m_Q^2 s}{t_1 u_1}-1\right)\,.
\end{equation}
Here, $t_1$ and $u_1$ are the usual tree-level Mandelstam variables.
Contrary to the unpolarized case,
$\Delta B_{\gamma g}$ receives no ${\cal{O}}(\varepsilon)$ contributions \cite{Bojak:1998bd}.
$e_Q$ is the electromagnetic charge of the heavy quark $Q$ in units of the coupling $e$, i.e., $e_c=2/3$,
and $e^2/4\pi=\alpha$.

All $2\rightarrow 3$ processes exhibit singularities related to collinear splittings 
off the incoming photon ($y \to +1$) and parton $b$
($y\to -1$) which need to be factorized into
the bare photon and proton PDFs at a scale $\mu_f$, respectively.
As for the soft contributions discussed above,
the kinematics collapses to the much simpler case of $2\rightarrow 2$ scattering,
such that compact analytical expressions, 
summarized by $d\Delta\hat{\sigma}^{(c\pm)}_{\gamma b}$
in Eq.~(\ref{eq:xsec-decomp}), can be obtained:
\begin{eqnarray}
\label{eq:collinear}
\nonumber
d\Delta\hat{\sigma}^{(c\pm)}_{\gamma b} &=& -(4\pi)^{\varepsilon-2} \Gamma[1+\varepsilon] 
\left(\frac{2}{\omega}\right)^{\varepsilon} \frac{s^{-1-\varepsilon}}{4\varepsilon} d{\mathrm{PS}}_2 
\\
\nonumber
&\times& \left[\left( \frac{1}{1-x} \right)_{\tilde{\rho}} -2\varepsilon 
\left( \frac{\log(1-x)}{1-x} \right)_{\tilde{\rho}} \right] \\
&\times& \Delta f_{\gamma b}^{(c\pm)}(s,m_Q,x,\theta_1)\,,
\end{eqnarray}
where
\begin{eqnarray}
\label{eq:coll1}
\Delta f_{\gamma g}^{(c+)}(x,\theta_1) 
&=& 0\,,\\
\nonumber
\Delta f_{\gamma g}^{(c-)}(x,\theta_1) 
&=& 32\pi \alpha_s s(1-x) \\
&\times& \Delta|M_{\gamma g}|^2\left|_{p_2\rightarrow xp_2} \right. \Delta P_{gg}(x)\,,\\
\nonumber
\Delta f_{\gamma q}^{(c+)}(x,\theta_1) 
&=& 32\pi \alpha e_Q^2 s(1-x) \\    
&\times &\Delta|M_{q\bar{q}}|^2\left|_{p_1\rightarrow xp_1}\right. \Delta P_{q\gamma}(x)\,,\\
\nonumber
\Delta f_{\gamma q}^{(c-)}(x,\theta_1) 
&=& 32\pi \alpha_s s(1-x) \\
\label{eq:coll4}
&\times& \Delta|M_{\gamma g}|^2\left|_{p_2\rightarrow xp_2}\right. \Delta P_{gq}(x)\,.
\end{eqnarray}
Here, $p_1$ and $p_2$ denote the momenta of the photon and parton $b$, respectively.
Both, the standard two-body phase-space $d{\mathrm{PS}}_2$ in (\ref{eq:collinear}) and
the Born matrix elements squared in Eqs.~(\ref{eq:coll1})-(\ref{eq:coll4})
are to be evaluated with appropriate collinear kinematics as indicated by the shift of
momenta $p_2\to x p_2$, etc; further details can be found in Refs.~\cite{Mangano:1991jk,Frixione:1993dg,Riedl:2009ye}.

The resulting $1/\varepsilon$ divergence in (\ref{eq:collinear})  
assumes the form dictated by the factorization theorem, i.e., a
convolution of $d$-dimensional helicity-dependent LO splitting functions
$\Delta P_{ij}(x)$ and Born matrix elements $\Delta|M_{ab}|^2$.
By adding appropriate counter cross sections to (\ref{eq:collinear}),
which to  ${\cal{O}}(\alpha \alpha_s^2)$ schematically read
\begin{eqnarray}
\label{eq:xsec-counter}
\nonumber
d\Delta\hat{\sigma}_{\gamma b}^{\tilde{c}}(\mu_f)
&=&-\frac{\alpha_s}{2\pi} \sum_i \int \frac{dx}{x} 
\Big[ \Delta {\cal{P}}_{i\gamma}(x,\mu_f) d\Delta\hat{\sigma}_{ib}^{(b)}(xs)\\
&+& \Delta {\cal{P}}_{ib}(x,\mu_f) d\Delta\hat{\sigma}_{\gamma i}^{(b)}(xs)\Big]\,,
\end{eqnarray}
where
\begin{equation}
\label{eq:trans-func}
{\cal{P}}_{ij}(x,\mu_f) = \Delta P_{ij}(x) [-\frac{1}{\varepsilon}+\gamma_E-\ln 4\pi 
+\ln\frac{\mu_f^2}{\mu^2}] + \Delta g_{ij}(x)\;,
\end{equation}
all collinear singularities can be consistently factorized into the scale evolution of
the bare photon and proton PDFs,
depending on whether they originate from collinear configurations 
involving the initial-state photon or parton $b$.
The required Born cross sections and LO $\Delta P_{ij}(x)$ are listed in
the Appendix of Ref.~\cite{Riedl:2009ye}, except for 
\begin{equation}
\label{eq:pqgamma}
\Delta P_{q\gamma} = C_A [2x-1-2\varepsilon (1-x)]\;.
\end{equation}
$C_A=3$ in (\ref{eq:pqgamma}) and, below, $C_F=4/3$ are the usual QCD color factors

The factorization scheme is fully specified by the choice of $\Delta g_{ij}$
in (\ref{eq:trans-func}) for which we take $\Delta g_{qq}=-4C_F (1-x)$, 
to guarantee helicity conservation within the 
HVBM prescription for $\gamma_5$ in $d$ dimensions \cite{ref:hvbm},
and $\Delta g_{ij}=0$ otherwise. The Euler constant 
$\gamma_E$ and $\ln 4\pi$, both, like the scale $\mu$, 
artifacts of dimensional regularization,
are subtracted along with the $1/\varepsilon$ singularity. 
This defines the $\overline{\mathrm{MS}}$ scheme in the polarized case, see, e.g., Ref.~\cite{ref:nlo-split}.

Finally, the last term in Eq.~(\ref{eq:xsec-decomp}), $d\Delta\hat{\sigma}^{(f)}_{\gamma b}$,
contains all the remaining, finite contributions, and the phase-space integration
can be performed numerically in four dimensions.
Monte Carlo integrations of (\ref{eq:xsec-fact})
for different measurement functions ${\cal{S}}$ can be done in parallel 
by randomly generating a sufficiently large sample
of final-state configurations characterized by
$x_1,x_2,x,y,\theta_1,$ and $\theta_2$ to account for the possibly large
cancellations among the various terms.
Remnants of the regularization of soft and collinear regions of
phase are the mathematical distributions appearing in the different contributions to 
Eq.~(\ref{eq:xsec-decomp}). Their proper definitions through a test function,
properties, and numerical treatment
are discussed at length in Refs.~\cite{Mangano:1991jk,Frixione:1993dg,Riedl:2009ye}
and need not be repeated here.

\section{Phenomenological Studies \label{sec:pheno}}
\subsection{Preliminaries\label{sec:prel}}
%
Based on our Monte Carlo code described above,
we pre\-sent a comprehensive study of charm quark photoproduction in 
longitudinally polarized muon-deuterium collisions
at a c.m.s.\ energy of $\sqrt{S}\simeq 18\,\mathrm{GeV}$
relevant for the COMPASS experiment at CERN.
Unless specified otherwise, we will show results for single-inclusive cross sections, 
differential in the transverse momentum $p_T^D$ of the observed $D$ meson.

The hadronization of the heavy quarks produced in the hard
scattering is modeled using Eq.~(\ref{eq:conv}) with \cite{ref:frag}
\begin{equation}
\label{eq:frag}
D^{Q\to H_Q}(z) = N_Q z^{\alpha_{Q}}(1-z)\;.
\end{equation}
The fragmentation function $D^{Q\to H_Q}$ is normalized such that its first moment
is unity, i.e., $N_Q = (\alpha_Q+1)(\alpha_Q+2)$.
For the remaining free parameter in (\ref{eq:frag}) we take
$\alpha_c=5$ as the default value in all our calculations;
see Table~4 in Ref.~\cite{ref:frag-review}.
To estimate the uncertainties associated with the choice of
$\alpha_c$, we will vary it in the range $3\le \alpha_c \le 7$ \cite{ref:frag-review}.
Like in the experimental analyses \cite{ref:compass-charm,ref:kurek,ref:franco-phd,ref:franco}, 
we adopt a cut $z>0.2$ throughout our phenomenological studies, 
where $z$ is the fraction of the energy of the photon
taken by the $D$ meson in the laboratory frame.

In the computation of the LO and NLO unpolarized cross sections we use the
LO and NLO CTEQ6 parton densities \cite{ref:cteq6} 
and values for the strong coupling $\alpha_s$, respectively.
In the polarized case, we adopt the best fit from the comprehensive global analysis performed by
the DSSV group \cite{deFlorian:2008mr}, which is the only set of helicity PDFs including constraints 
on $\Delta g(x,\mu_f^2)$ from $pp$ data. The resulting PDFs are characterized by a small $\Delta g(x,\mu_f^2)$ in the 
$x$-range predominantly probed by RHIC experiments, 
$\int_{0.05}^{0.2} \Delta g(x,10\,\mathrm{GeV}^2) dx = 0.005^{+0.129}_{-0.164}$,
with a node at $x\simeq 0.1$ \cite{deFlorian:2008mr}.
To study the sensitivity of charm photoproduction data to
different $\Delta g(x,\mu_f)$, we also use the GRSV ``standard'' set \cite{ref:grsv},
which has a positive $\Delta g$ that is larger in size than the one of DSSV.
We note that the most recent RHIC data \cite{ref:starnew,ref:rhicwp} tend to prefer a polarized gluon
density somewhere in between the ones obtained in the DSSV and GRSV fits \cite{ref:dssvplus,ref:rhicwp}.

Equation~(\ref{eq:xsec-fact}) consistently includes both direct
and resolved photon contributions to the spin-dependent 
photoproduction cross section
by defining $\Delta f^l_a$ as the convolution of the 
polarized lepton-to-photon splitting function
$\Delta P_{\gamma l}$ and the structure functions $\Delta f_a^{\gamma}$
of a circularly polarized photon, i.e.,
\begin{equation}
\Delta f^l_a(x_a,\mu_f)=
\int_{x_a}^1\frac{dy}{y}
\Delta P_{\gamma l}(y)
\Delta f^{\gamma}_a\left(\!x_{\gamma}=
\frac{x_a}{y},\mu_f\!\right).
\label{eq:leptonpdf}
\end{equation}
Nothing is known experimentally about the $\Delta f_a^{\gamma}$, which account for
the hadron\-ic structure of a photon in the resolved case where $a\neq \gamma$ in (\ref{eq:xsec-fact}).
However, as will be demonstrated below, uncertainties associated with
$\Delta f_a^{\gamma}$ turn out to be negligible for charmed meson photoproduction at COMPASS
kinematics. Contrary to inclusive hadron production \cite{Jager:2003vy},
the resolved contribution turns out to be 
numerically small even when estimated with the ``maximum'' model for $\Delta f_a^{\gamma}$
introduced in Ref.~\cite{ref:polphoton} and based on saturating the positivity limit 
$|\Delta f_a^{\gamma}(x,\mu_f)|\le f_a^{\gamma}(x,\mu_f)$ at some
low scale $\mu_f$ using the unpolarized set of GRV photon distributions
\cite{ref:grvphoton} as reference.
The direct part of the cross section (\ref{eq:xsec-fact}) where $a=\gamma$
is obtained by setting
\begin{equation}
\Delta f^{\gamma}_a(x,\mu_f)=\delta(1-x)\;.
\label{eq:direct}
\end{equation}
in Eq.~(\ref{eq:leptonpdf}). 

The collinear emission of a quasi-real photon with low virtuality $Q$
and momentum fraction $y$ off a muon with mass $m_{\mu}$ 
is given by the Weizs\"acker-Williams equivalent photon spectrum
which reads in the polarized case \cite{ref:polww}
\begin{eqnarray}
\Delta P_{\gamma l}(y)&=&
\frac{\alpha}{2\pi}
\left[\frac{1-(1-y)^2}{y}\ln{\frac{Q_{\mathrm{max}}^2(1-y)}{m_{\mu}^2y^2}}\right]\nonumber\\
&+&\left. 2m_{\mu}^2y^2\left(\frac{1}{Q_{\mathrm{max}}^2}
-\frac{1-y}{m_{\mu}^2y^2}\right)\right] \, .
\label{eq:polww}
\end{eqnarray}
The upper limit $Q_\mathrm{max}$ is determined by experimental conditions.
For COMPASS we take $Q^2_{\mathrm{max}}=0.5\,\mathrm{GeV}^2$ and, in addition,
restrict the fraction $y$ of the muon's momentum taken by the quasi-real photon
to the range $0.1\le y \le 0.9$.

We use $m_{c}=1.35\,\mathrm{GeV}$ as the value of the charm quark mass
for all our results.
For the factorization and renormalization scales in Eq.~(\ref{eq:xsec-fact})
we take $\mu_f=\mu_r=\xi(p_T^2+m_c^2)^{1/2}$ 
with $\xi=1$ as the central value and where $p_T$ denotes the transverse momentum
of the charm quark. As is commonly done, we vary them simultaneously in the
range $1/2\le \xi\le 2$ to estimate the residual scale dependence at NLO which
represents the dominant source of theoretical uncertainty and
can be taken as a rough measure of yet unknown higher order corrections.

\begin{figure*}[bht!]
\vspace*{-0.35cm}
\begin{center}
\includegraphics[width=0.435\textwidth]{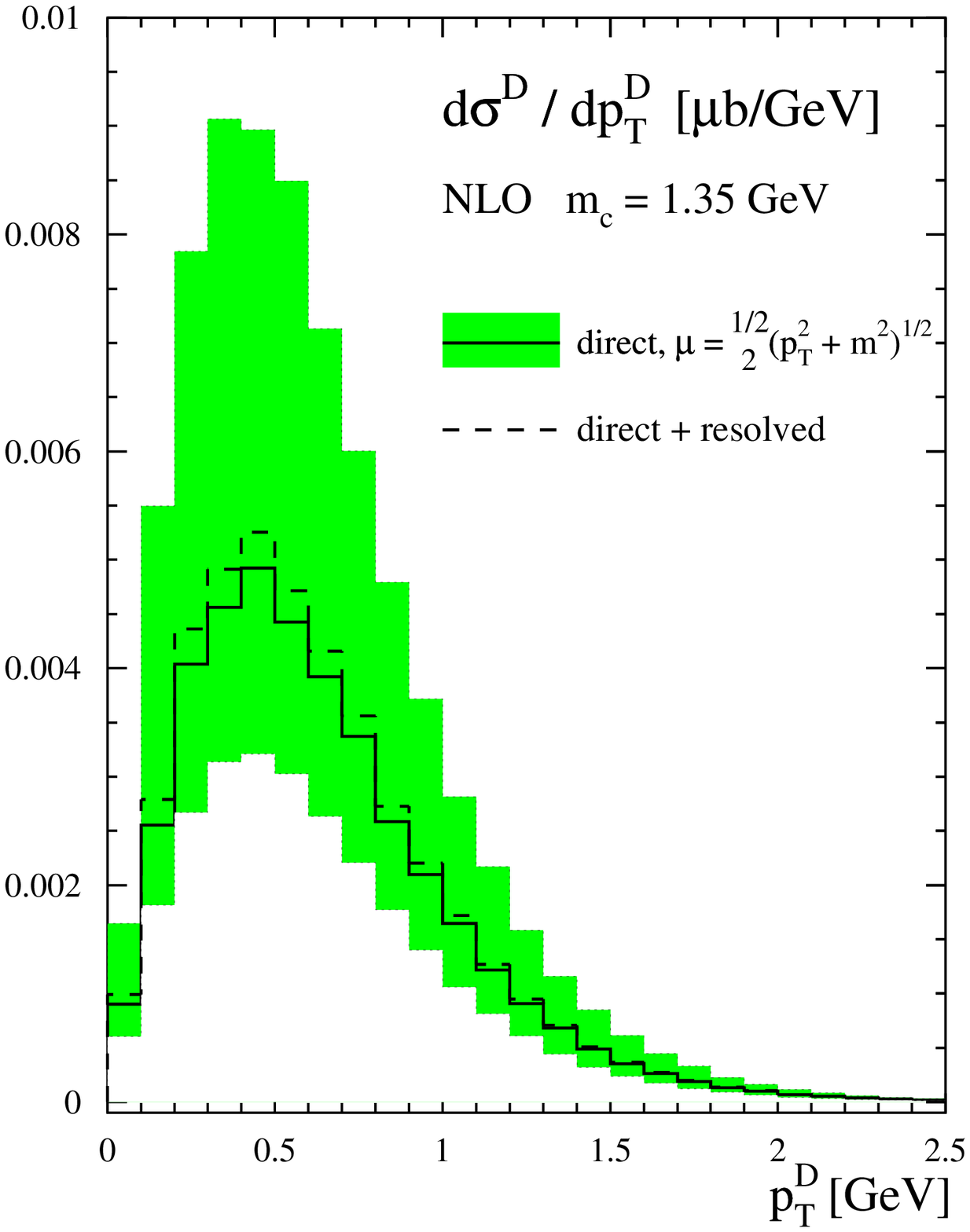}
\includegraphics[width=0.435\textwidth]{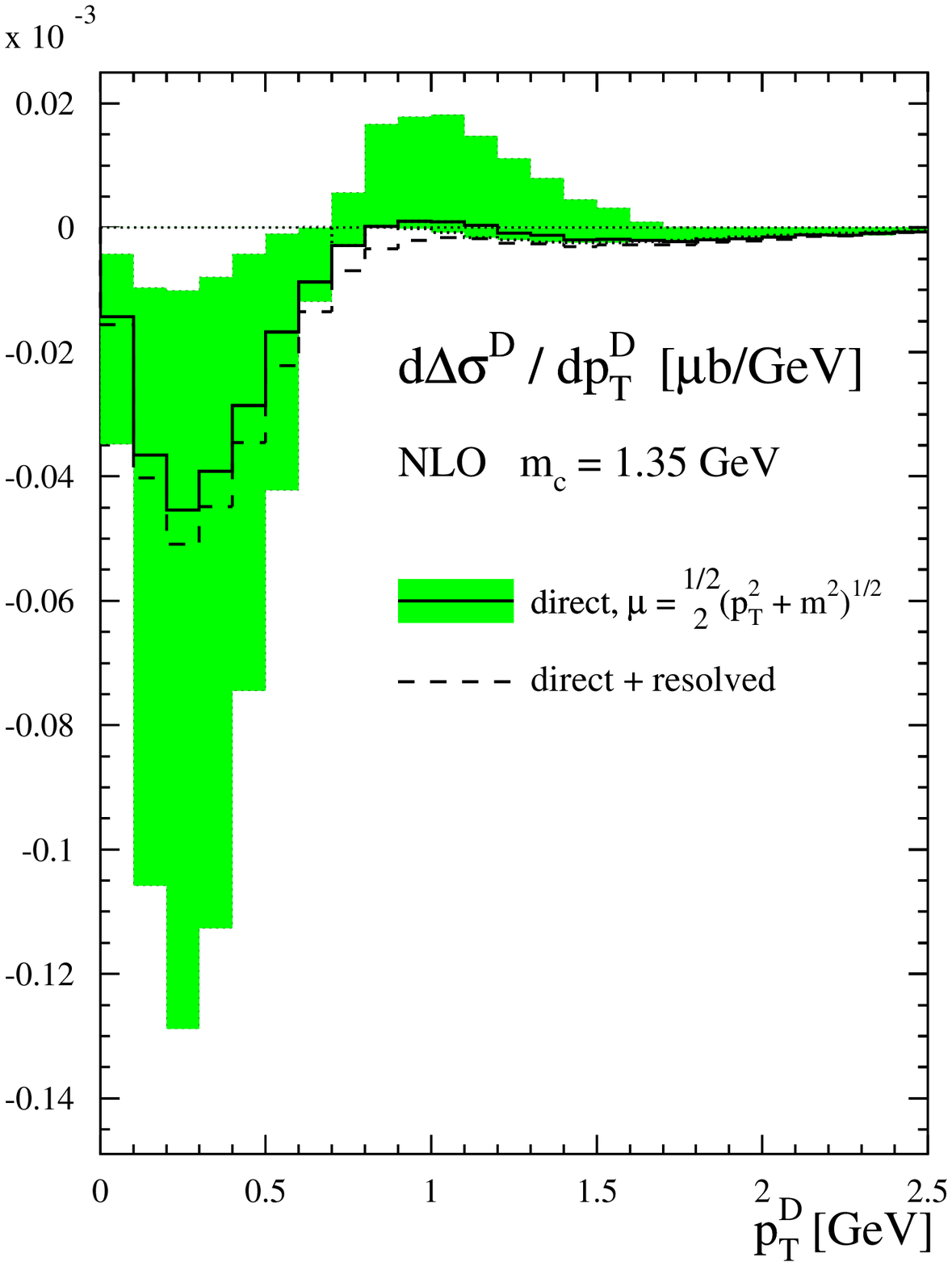}
\end{center}
\vspace*{-0.35cm}
\caption{\label{fig:figure2}
Scale dependence of the single-inclusive transverse momentum spectrum
of $D$ mesons at NLO with $E_D<30\,\mathrm{GeV}$ for the dominant direct photon contribution
in unpolarized ({\bf left-hand side}) and
polarized ({\bf right-hand side}) muon-deuterium collisions at COMPASS
kinematics; see text. 
Factorization and renormalization scales are varied simultaneously
in the range $\mu=\xi(p_T^2+m_c^2)^{1/2}, 1/2\le\xi\le 2$; the solid line
refers to $\xi=1$.
In both panels, the dashed line shows the result for the sum of
direct and resolved photon cross sections for $\xi=1$.}
\end{figure*}
%
\subsection{Numerical Results\label{sec:results}}
%
We begin our detailed numerical studies with a discussion of 
the scale dependence of the
unpolarized and polarized photoproduction cross sections for
$D$ mesons in muon-deuterium collisions at COMPASS.
Figure~\ref{fig:figure2} shows the scale ambiguity from
varying $\mu_{f,r}=\xi(p_T^2+m_c^2)^{1/2}$ simultaneously in the
range $1/2\le\xi\le 2$ for the dominant direct photon contribution ($a=\gamma$) only. 
The solid lines denote our default
choice of scales, $\xi=1$. Since the published COMPASS data \cite{ref:compass-charm} 
are divided into three bins in the energy $E_D$ of the produced $D$ meson,
$E_D <30\,\mathrm{GeV}$, $30\le E_D\le 50\,\mathrm{GeV}$, and
$E_D>50\,\mathrm{GeV}$, we impose the cut $E_D<30\,\mathrm{GeV}$ in Fig.~\ref{fig:figure2}.
Results for the other two bins in $E_D$ are very similar and hence not shown.

\begin{figure}[h!]
\vspace*{-0.35cm}
\begin{center}
\includegraphics[width=0.425\textwidth]{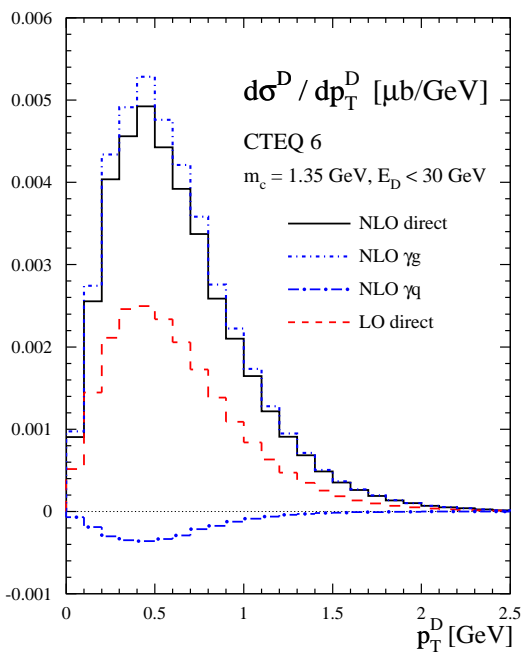}
\end{center}
\vspace*{-0.35cm}
\caption{\label{fig:figure3}
Contributions of the individual $\gamma g$ and $\gamma q$ initiated 
subprocesses to the direct photoproduction cross section at NLO for $\xi=1$
and $E_D<30\,\mathrm{GeV}$ (solid line). 
Also shown is the LO result (dashed line).} 
\end{figure}
As can be seen, the theoretical uncertainties due to the choice of scale
are quite sizable, both in the unpolarized and in the polarized case
and can be even further inflated by varying $\mu_f$ and $\mu_r$ independently
as was done, for instance, in our study of heavy quark hadroproduction \cite{Riedl:2009ye}.
Also, possible variations of $m_c$, which we do not pursue here, 
would add to the theoretical error. 
By comparing the solid and dashed lines in Fig.~\ref{fig:figure2}
one can infer the relevance of the resolved photon cross section ($a\neq \gamma$).
As was argued in Sec.~\ref{sec:prel}, its contribution is rather small 
compared to the direct photon cross section and potential
uncertainties due to the unknown $\Delta f^{\gamma}_a$ do not matter.

Next, we study the role of the individual partonic subprocesses and their contribution
to the photoproduction cross section in Eq.~(\ref{eq:xsec-fact}).
Figure~\ref{fig:figure3} displays the decomposition of the unpolarized direct photon
cross section for $\xi=1$, as shown in Fig.~\ref{fig:figure2}, in terms of the $\gamma g$ and
$\gamma q$ processes.
Again, we choose $E_D<30\,\mathrm{GeV}$, and results for the other two bins in $E_D$ are
very similar.
As expected, the photon-gluon fusion mechanism, which is already present at LO, 
gives the main contribution to the cross section. The genuine NLO photon-quark channel
yields a negative but small correction. 
The NLO result for the $D$ meson production cross section is roughly a factor of two larger than
the corresponding estimate at LO as can be gathered from comparing the dashed and solid lines
in Fig.~\ref{fig:figure3}. 

%
\begin{figure*}[th]
\vspace*{-0.35cm}
\begin{center}
\includegraphics[width=0.435\textwidth]{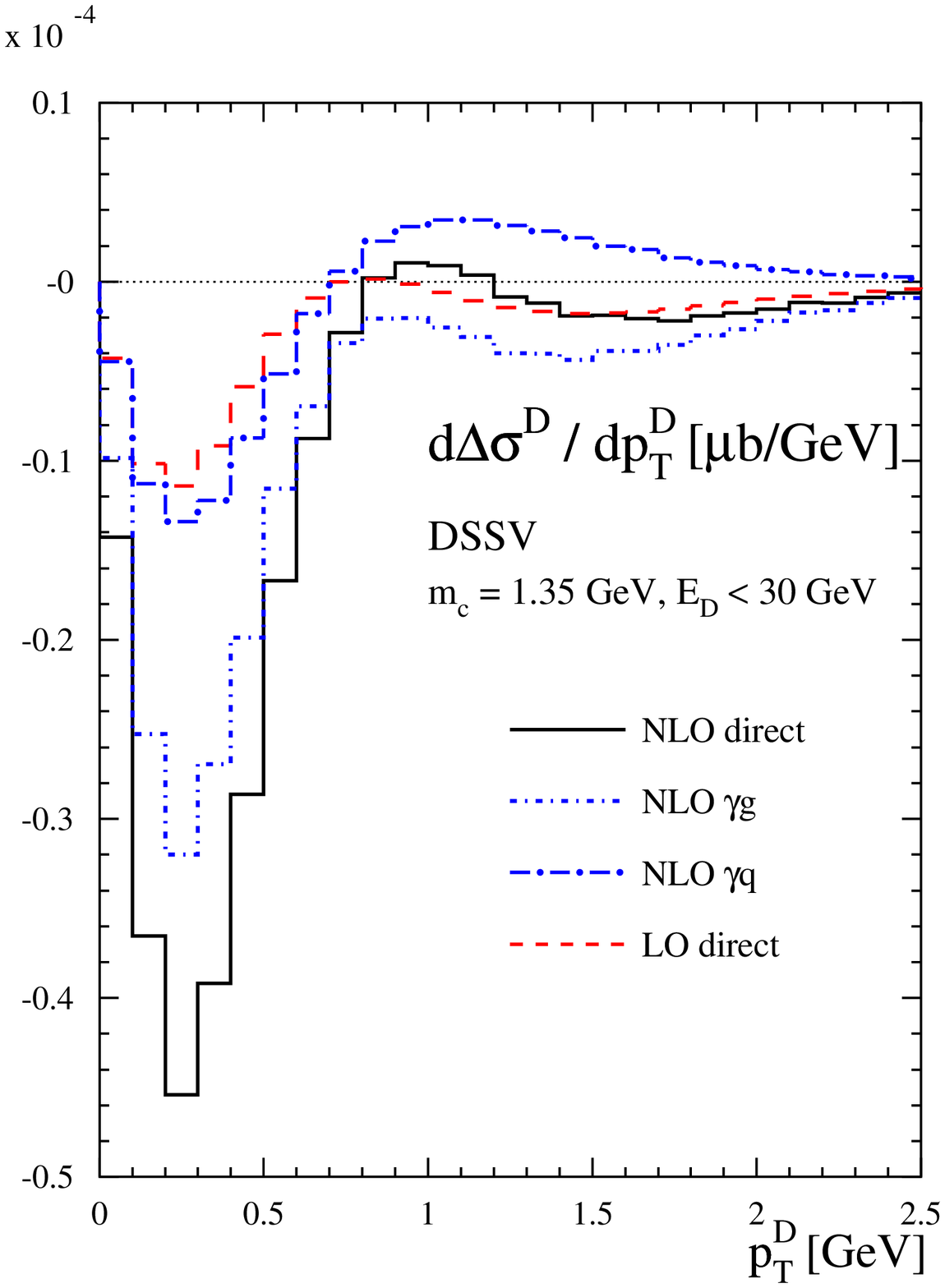}
\includegraphics[width=0.435\textwidth]{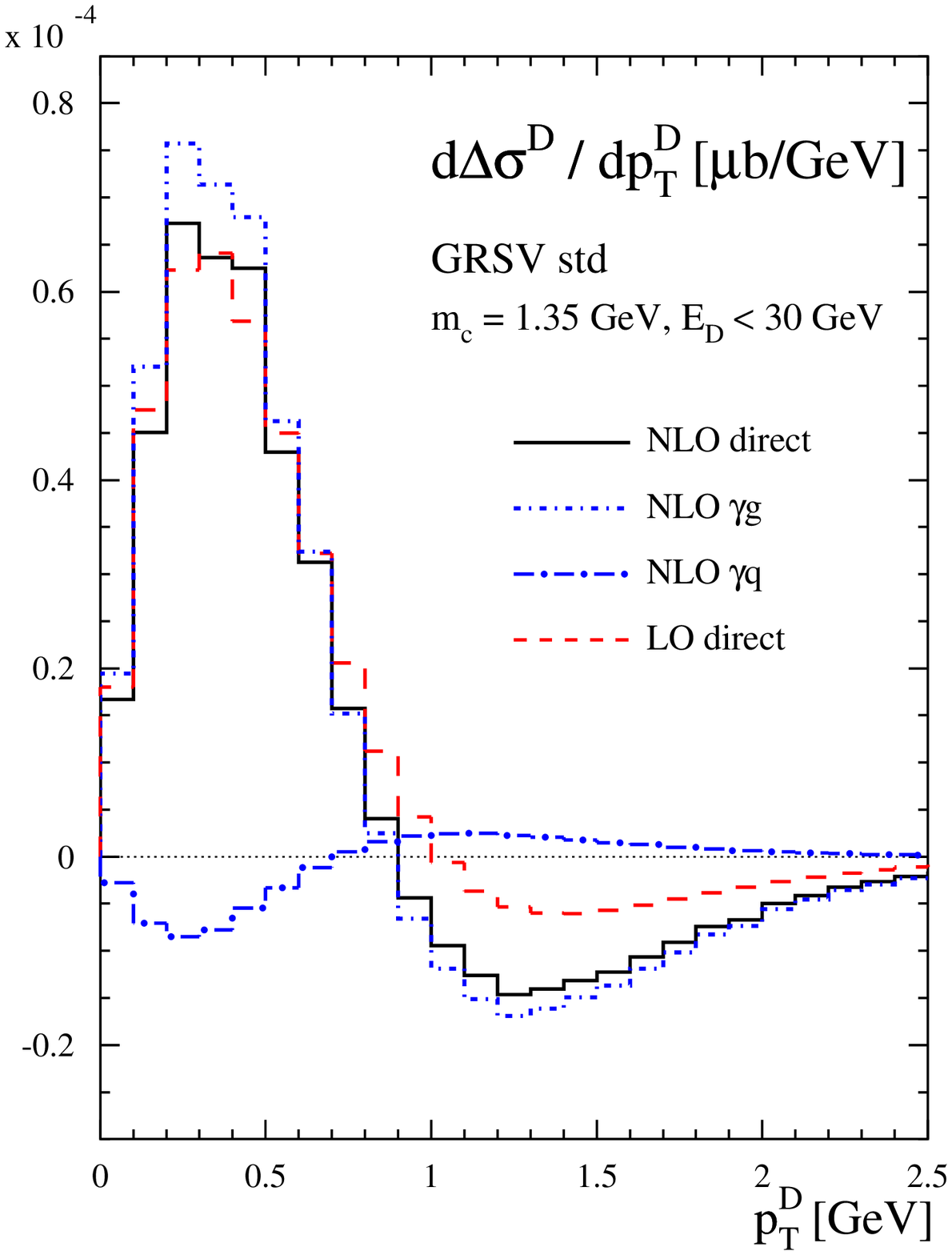}
\end{center}
\vspace*{-0.35cm}
\caption{\label{fig:figure4}
As in Fig.~\ref{fig:figure3} but now for the spin-dependent photoproduction
cross section for two different choices of helicity PDFs:
DSSV \cite{deFlorian:2008mr}  ({\bf left-hand side}) 
and GRSV ``standard'' \cite{ref:grsv} ({\bf right-hand side}).}
\end{figure*}
A similar exercise in the polarized case is shown in Fig.~\ref{fig:figure4}
for two different choices of helicity PDFs.
First, one notices that the results obtained with the DSSV and GRSV sets
differ in sign and magnitude of the cross section, which is readily explained 
by the very different gluon densities in both sets.
The positive definite $\Delta g(x,\mu_f)$ of GRSV leads to a similar decomposition
into $\gamma g$ and $\gamma q$ subprocesses as was observed in the unpolarized 
case in Fig.~\ref{fig:figure3}. Again, the cross section is strongly dominated
by photon-gluon fusion, and the $\gamma q$ channel only yields a small correction.
On the contrary, the oscillating $\Delta g(x,\mu_f)$ of the DSSV set of helicity PDFs
leads to a negative $d\Delta\sigma^D$. 
Since the DSSV gluon is much smaller in size than the one of GRSV, the genuine NLO 
photon-quark contribution, which is numerically very similar in both PDF sets, is more
important and yields more than a quarter of the cross section at small $p_T^D$.
Another important observation concerns the relevance of NLO corrections
which appears to be very different for the DSSV and GRSV helicity PDFs. This implies
that higher order QCD effects do {\em not} cancel in the experimentally relevant
double-spin asymmetry, 
\begin{equation}
\label{eq:all}
A_{LL}\equiv \frac{d\Delta\sigma^D}{d\sigma^D}\,\,,
\end{equation}
which we shall discuss in more detail below.
Clearly, for a reliable quantitative analysis of charm photoproduction in terms
of polarized PDFs, preferably as part of a global QCD fit, 
it is indispensable to properly include NLO corrections.

%
\begin{figure*}[th!]
\vspace*{-0.35cm}
\begin{center}
\includegraphics[width=0.425\textwidth]{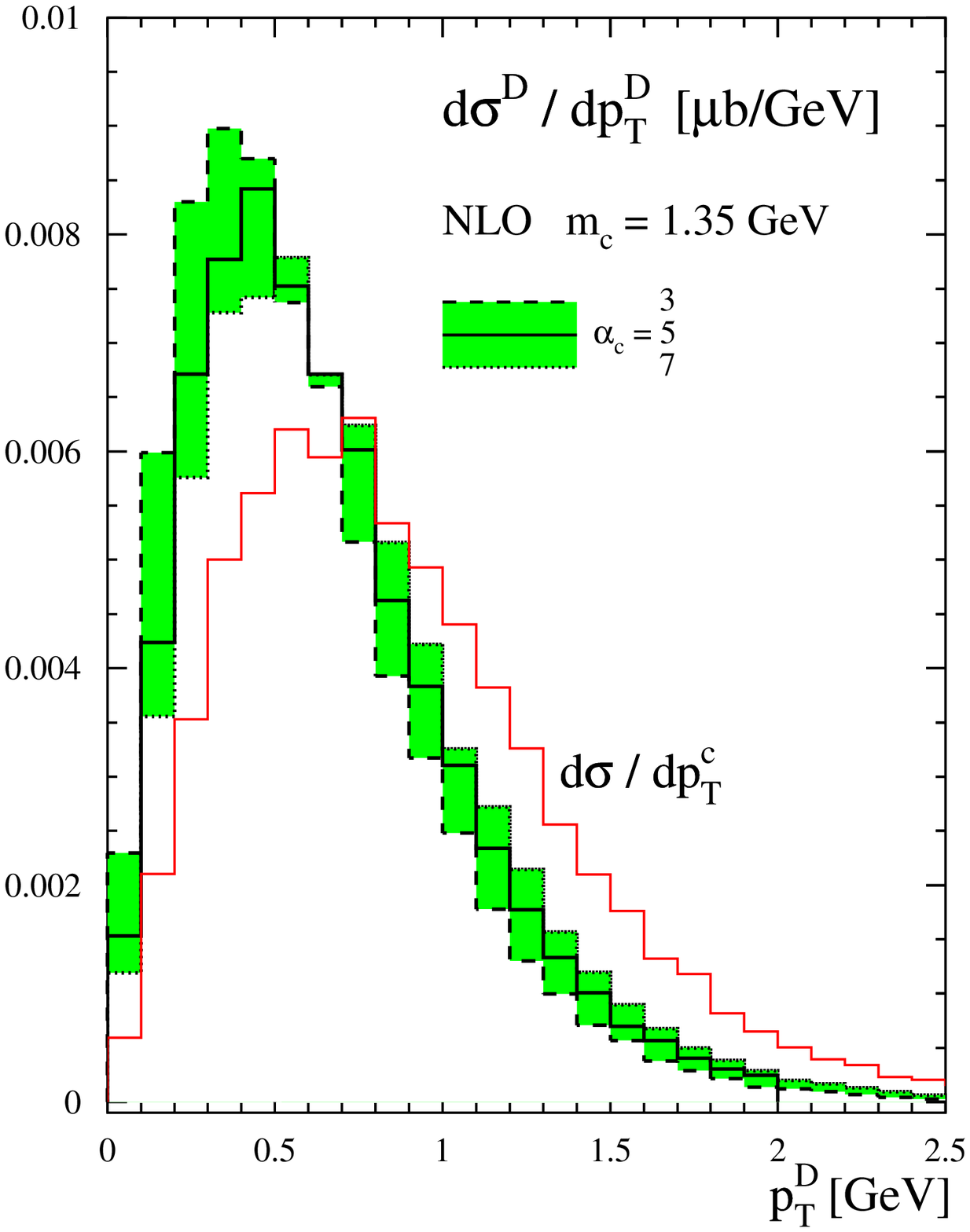}
\includegraphics[width=0.425\textwidth]{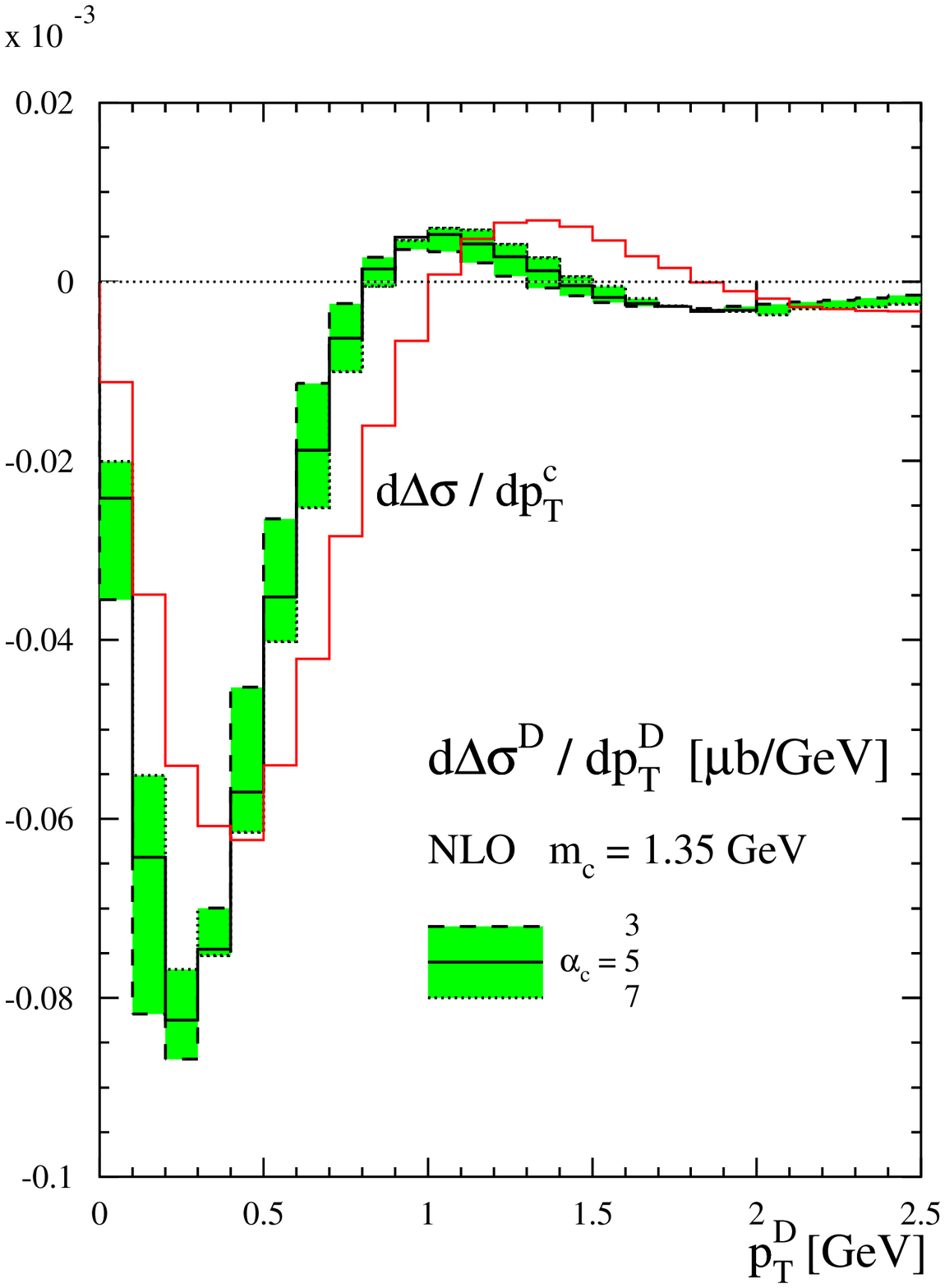}
\end{center}
\vspace*{-0.35cm}
\caption{\label{fig:figure5}
As in Fig.~\ref{fig:figure2}, but now for variations of the parameter
$\alpha_c$ in the fragmentation function (\ref{eq:frag}) 
for the direct cross section computed with $\xi=1$.
No cuts on the $D$ meson or charm quark are imposed; see text.
Also shown are the results on the charm quark level, $d(\Delta)\sigma/dp_T^c$.}
\vspace*{-0.35cm}
\begin{center}
\includegraphics[width=0.425\textwidth]{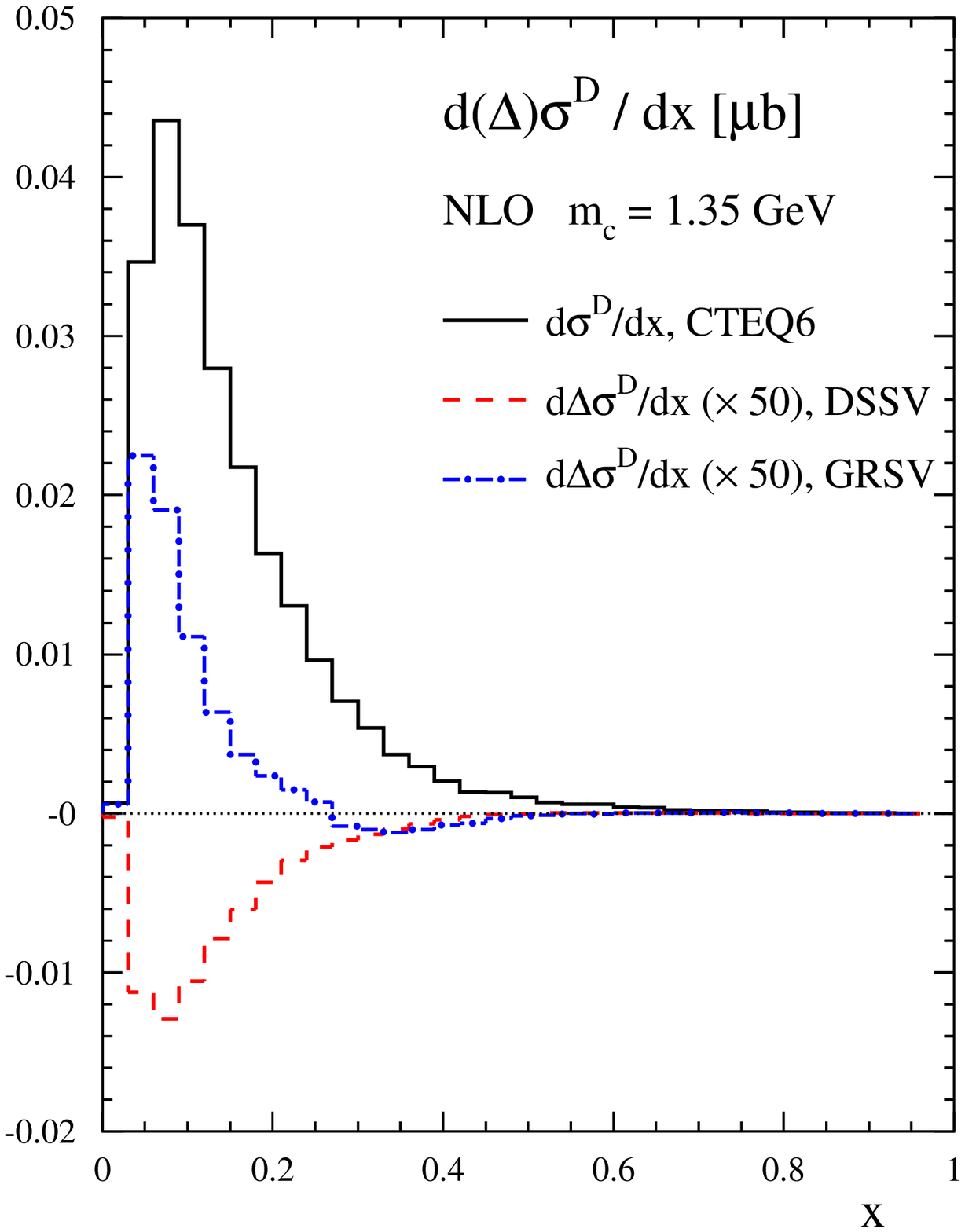}
\includegraphics[width=0.425\textwidth]{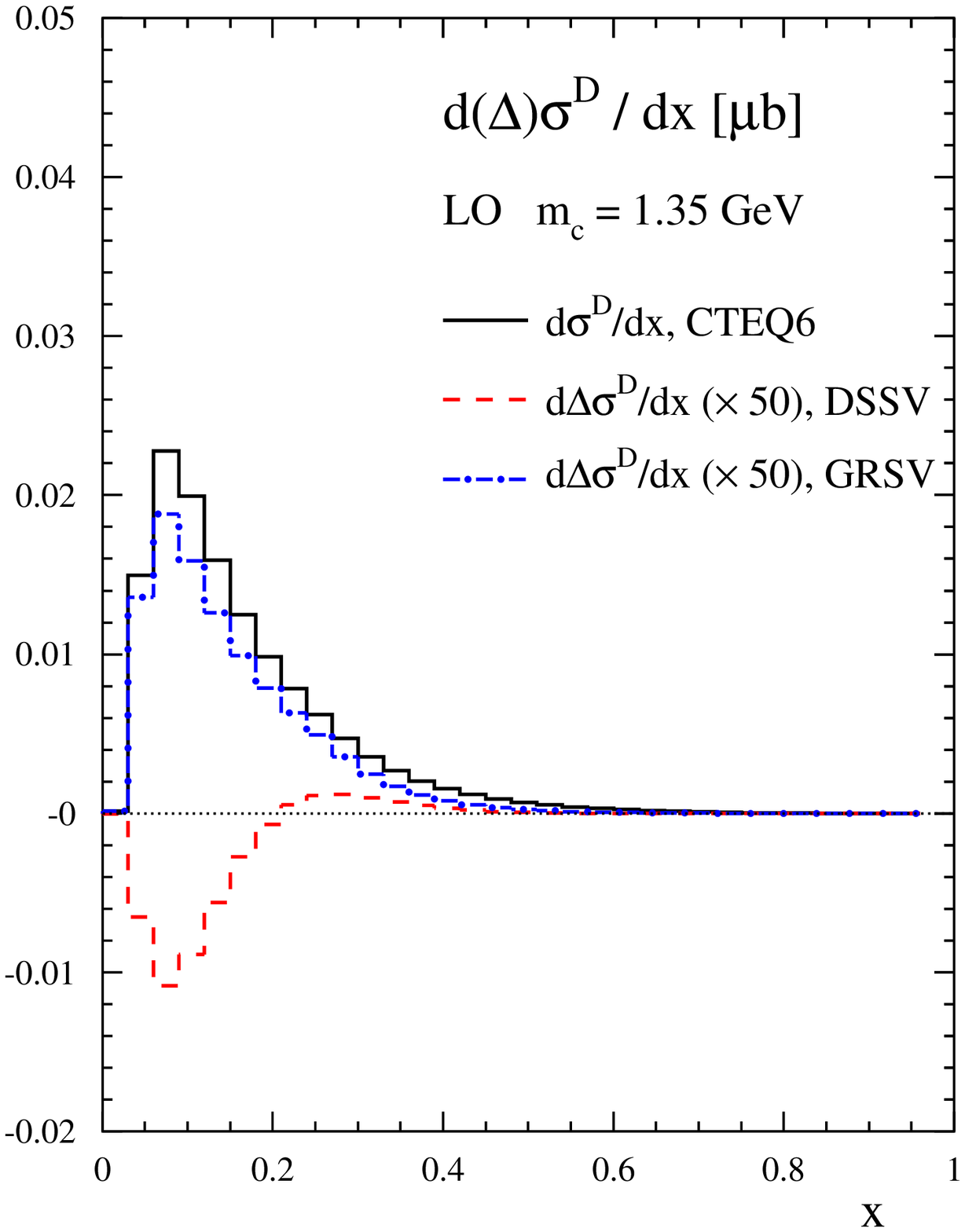}
\end{center}
\vspace*{-0.35cm}
\caption{\label{fig:figure6}
Distribution $d(\Delta)\sigma^D/dx$ in the momentum fraction $x$ probed in the PDFs
at NLO ({\bf left-hand side}) and LO ({\bf right-hand side}) accuracy
for both the unpolarized and polarized direct photon contribution,
integrated over phase-space and without imposing cuts. Note that the polarized results are
scaled by a factor of 50 for better visibility.}
\vspace*{-0.25cm}
\end{figure*}
Figure~\ref{fig:figure5} illustrates the uncertainties due to our choice
of $\alpha_c$ in the fragmentation of the charm quark into the observed
$D$ meson.
As in Fig.~\ref{fig:figure2}, we show the direct photon contribution for
$\xi=1$, but here we do not impose any cuts on the $D$ meson to allow for a
better comparability with the cross sections on the charm quark level
which are presented as well.
Compared to the factorization scale ambiguities, the dependence of
our results on $\alpha_c$ is fairly moderate for both the unpolarized
and the polarized cross sections. A similar observation was
made in the case of hadroproduction \cite{Riedl:2009ye}.
Since charm quarks lose only very little of their momentum 
during hadronization, i.e., $D^{Q\to H_Q}(z)$ is peak\-ed at fairly large values of $z$,
the convolution (\ref{eq:conv}) introduces only a rather small shift 
in the transverse momentum spectrum of the charm quarks. This can be inferred 
from the curves labeled $d(\Delta)\sigma/dp_T^c$ in Fig.~\ref{fig:figure5}.

Another interesting question concerns the range of momentum fractions $x$ predominantly probed in
the PDFs by the COMPASS data. Due to the dominance of the photon-gluon fusion process,
charm photoproduction will mainly lead to a constraint on the
gluon helicity distribution, which is the prime motivation for such measurements. 
The $x$ distribution in LO and NLO,
for both the unpolarized and polarized direct photon contribution to
the cross section (\ref{eq:xsec-fact}) is explored in Fig.~\ref{fig:figure6}.
No cuts are imposed in this calculation and the $D$ meson spectrum
is integrated over the entire phase space.
As has to be expected from the fact that the PDFs enter
the cross section (\ref{eq:xsec-fact}) through a convolution, 
a broad range of $x$ values is sampled. It turns out, however,
that the mean value of $x$, where the distribution $d(\Delta)\sigma^D/dx$
is peaked, is fairly independent not only of the order in perturbation
theory, LO or NLO, but to a large extent also of the chosen set of polarized or unpolarized PDFs.
We roughly estimate the average momentum fraction to be 
$\langle x\rangle \simeq 0.08$ with an error of about $^{+0.12}_{-0.03}$.

Our results differ from preliminary estimates of $\langle x \rangle$ 
by the COMPASS collaboration \cite{ref:kurek,ref:franco-phd,ref:franco}, 
where NLO results have been obtained based on some parton shower Monte Carlo to approximate 
the phase space for the NLO matrix elements of Ref.~\cite{Bojak:1998bd}. Significant differences between
$\langle x \rangle$ estimated in LO and NLO are found in this way. While their LO result
for $\langle x \rangle$ agrees with our estimate of about 0.08, their preliminary NLO result is 
$\langle x \rangle = 0.28_{-0.10}^{+0.19}$. Since the details of the
method are not yet published, it is not yet clear how these results can be compared
to our full NLO calculation.
We also note that once data on photoproduction processes are implemented in global QCD
analyses of helicity PDFs, information on $\langle x\rangle$, though useful,
is no longer required or relevant as the fits automatically impose the constraints from data 
for any given functional form assumed for the $\Delta f(x,\mu_f)$.

%
%
\begin{figure}[th]
\vspace*{-0.35cm}
\begin{center}
\includegraphics[width=0.45\textwidth]{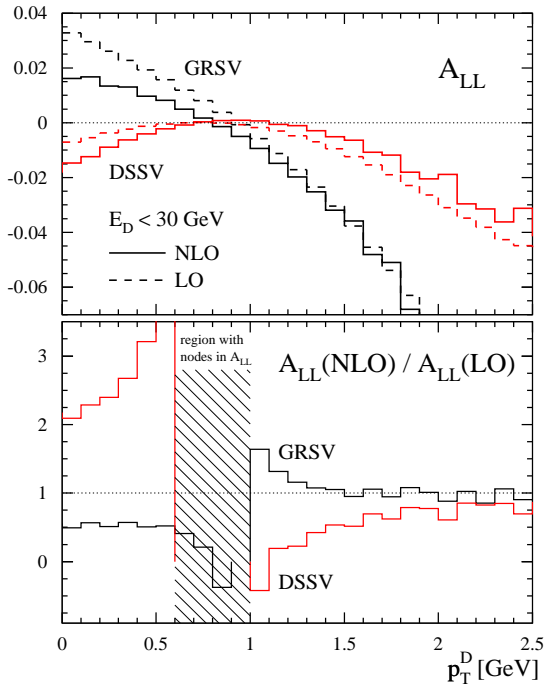}
\end{center}
\vspace*{-0.35cm}
\caption{\label{fig:figure7}
{\bf Upper panel}: double spin asymmetry in LO and NLO for
$D$ meson photoproduction at COMPASS kinematics for
the DSSV and GRSV sets of helicity PDFs; as in Figs.~\ref{fig:figure2}
and \ref{fig:figure3} a cut $E_D<30\,\mathrm{GeV}$ is imposed;
{\bf lower panel}: the corresponding ratio of the NLO and LO results.
The region where $A_{LL}$ changes sign and the ratio becomes
meaningless is indicated by the shaded band.}
\end{figure}
Next, we turn to the experimentally measured double-spin asymmetry, defined in
Eq.~(\ref{eq:all}), which was analyzed at LO accuracy and under certain simplifying assumptions
in terms of the mean gluon polarization $\Delta g(\langle x\rangle,\mu_f)/$\linebreak $g(\langle x\rangle,\mu_f)$ by the 
COMPASS collaboration \cite{ref:compass-charm}; see the discussion below. Preliminary NLO estimates
are also available at NLO with the hybrid method outlined above 
\cite{ref:kurek,ref:franco-phd,ref:franco}.
In Fig.~\ref{fig:figure7} we show $A_{LL}$ for the two sets
of helicity PDFs used throughout this paper, 
by computing the ratio of the cross sections shown in Figs.~\ref{fig:figure2}
and \ref{fig:figure3} for $E_D<30\,\mathrm{GeV}$.
To resolve the differences in $A_{LL}$ obtained with the DSSV and GRSV PDFs, which mainly stem from
$\Delta g$, an experimental precision of at least $\delta A_{LL}\simeq 0.02$ needs to be achieved.

We also compare NLO and LO estimates of $A_{LL}$ for both sets of helicity PDFs 
in the lower panel of Fig.~\ref{fig:figure7} and
find rather different patterns depending on $p_T^D$.
At small $p_T^D$, the NLO $A_{LL}$ is about a factor of two larger than the LO estimate 
for the DSSV set whereas a reduction by roughly the same amount is found with the
GRSV PDFs. This illustrates that any approximations for
the spin asymmetry, either to assume a cancellation of NLO corrections or a constant
pattern independent of the choice of PDFs,
are not justified and should not be used for analyzing data.
Again, only a global analysis will lead to consistent results. 
Qualitatively very similar results have been obtained for other cuts on the energy $E_D$ of the
observed $D$ meson, $30\le E_D \le 50\,\mathrm{GeV}$ and $E_D>50\,\mathrm{GeV}$.

%
%
\begin{figure}[th]
\vspace*{-0.35cm}
\begin{center}
\includegraphics[width=0.45\textwidth]{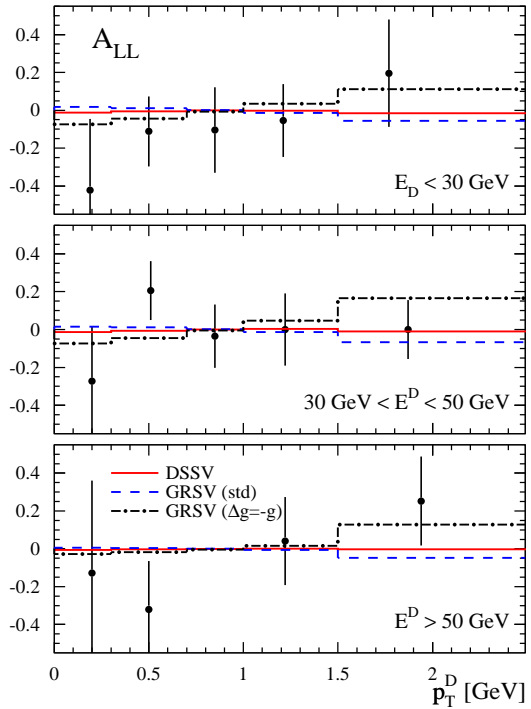}
\end{center}
\vspace*{-0.35cm}
\caption{\label{fig:figure8}
Predictions for the double-spin asymmetry for $D$ meson photoproduction
at COMPASS at NLO accuracy for three different sets of helicity PDFs
and in three bins of $E_D$ compared to data.
Note that in the bottom panel one of the data points is outside the
range shown for $A_{LL}$.}
\end{figure}
Finally, we compare our calculations at NLO accuracy with the available results from the 
COMPASS experiment.
Figure~\ref{fig:figure8} shows the data in three bins of the
energy $E_D$ of the detected $D$ meson as a function of its transverse momentum
$p_T^D$, imposing the cut $z>0.2$. 
Note that instead of using the published data \cite{ref:compass-charm}, we show new, 
preliminary results presented recently in Ref.~\cite{ref:kurek,ref:franco-phd,ref:franco}. 
A weighted average is performed
to combine the results for the three decay channels $D\to K\pi$, $D\to K\pi\pi$, and
$D\to K\pi\pi\pi$ listed in \cite{ref:kurek,ref:franco-phd,ref:franco}. 
We adopt the photon polarization dilution factors tabulated in Ref.~\cite{ref:kurek,ref:franco-phd,ref:franco}
to convert the data for $A_{LL}$ given in the photon-nucleon system to the
double-spin asymmetries for muon-nucleon scattering computed with our Monte Carlo code.

Clearly, present experimental uncertainties are too large to discriminate between different 
spin-dependent gluon densities. Apart from the DSSV and GRSV ``standard'' sets, we
also use an extreme GRSV set, characterized by a very large and negative
gluon density based on setting $\Delta g(x,\mu_0)=-g(x,\mu_0)$ at 
some low bound-state like
input scale $\mu_0$ \cite{ref:grsv} for the evolution. 
Despite leading to a distinctively different $A_{LL}$ than our two default sets,
all results are compatible with data within the experimental precision.

To overcome the statistical limitations, fewer bins have to be used or all data need
to be combined. COMPASS has performed such an analysis \cite{ref:compass-charm} yielding the result for
the gluon polarization quoted in Fig.~\ref{fig:deltag}, which is compatible 
with similar extractions of $\Delta g/g$ from hadron production data.
These kind of analyses are LO estimates, assuming, in addition, that
the convolutions of the PDFs with the partonic hard cross sections
in Eq.~(\ref{eq:xsec-fact}) can be approximated as
\begin{equation}
A_{LL} = \frac{\Delta g \otimes  d\Delta \hat{\sigma}_{\gamma g}}
{g \otimes d\Delta \hat{\sigma}_{\gamma g}} \approx \frac{\Delta g}{g}
\Big\langle \frac{d\Delta\hat{\sigma}_{\gamma g}}{d\hat{\sigma}_{\gamma g}} \Big\rangle\;.
\end{equation}
Here, the ``analyzing power'' $\langle \frac{d\Delta\hat{\sigma}_{\gamma g}}{d\hat{\sigma}_{\gamma g}}\rangle $
is evaluated at some average kinematics.
As mentioned before, in a proper glo\-bal analysis of helicity PDFs one does not need to resort
to any of these approximations whose validity is difficult to estimate or justify.
With our new Monte Carlo program it is now possible to perform such an analysis 
at NLO accuracy in the future. As we have demonstrated in some detail, NLO
corrections are indispensable for a quantitative analysis. Most importantly,
they do not cancel in the spin asymmetry as one might naively expect.

Apart from the phenomenological applications pre\-sen\-ted here, our code can
be also used to estimate spin asymmetries for charm and bottom photoproduction
and their impact on our knowledge of the spin structure of the nucleon
at higher c.m.s.\ energies intended for a first polarized lepton-ion collider.
Details on the EIC project can be found in \cite{Boer:2011fh}.
Such a machine will be indispensable to finally unravel the quark and gluon
contributions to the nucleon spin \cite{Aschenauer:2012ve}.

\section{Summary and Outlook}
%
We have presented a flexible parton-level Monte Carlo program to
compute heavy flavor distributions at NLO accuracy 
in longitudinally polarized lepton-nucleon collisions in the
photoproduction regime. For the first time, 
we consistently include both direct and resolved photon contributions
and found the latter to be negligibly small for charm production
at COMPASS kinematics.
Experimental acceptance cuts, the hadronization of the produced heavy quark pair,
and, if needed, their subsequent semi-leptonic decays can be included in phenomenological
applications.

Heavy flavor photoproduction receives its importance for the field of spin physics 
from its expected strong sensitivity to the polarized gluon density which we
confirm. In general, higher order corrections are found to be sizable and strongly dependent
on the chosen set of helicity PDFs. This is also true for the experimentally
relevant spin asymmetry despite naive expectations that QCD corrections
cancel in the ratio. Theoretical uncertainties due to the choice of the
factorization scale are sizable even at next-to-leading order accuracy
while ambiguities from the exact form of the charm quark fragmentation function
are less important.

The results obtained in this paper allow one to include data
on charm photoproduction consistently into future global analyses
of helicity PDFs. We have shown that data from COMPASS can lead in principle 
to a constraint on the polarized gluon density at a momentum
fraction of about 0.1.
Currently available data will have, however, very little impact on 
existing fits due to the size of the experimental uncertainties
which are too large to discriminate between different gluon densities.

In addition to the phenomenological studies performed in this paper, 
our code will be useful in assessing the physics impact of heavy quark
photoproduction at a possible future polarized lepton-ion collider like
the EIC project.

\section*{Note Added}
While completing our analysis, the preliminary COMPASS
results \cite{ref:kurek,ref:franco-phd,ref:franco} discussed in this paper 
have been published in \cite{ref:compass-new}. Our conclusions remain
unchanged.

\section*{Acknowledgements}
J.R.\ was supported by a grant of the ``Cusanuswerk'', Bonn, Germany.
M.S.\ acknowledges support by the U.S.\ Department of Energy (DOE) under contract 
No.~DE-AC02-98CH10886 and, in part, by a BNL ``Laboratory Directed Research and
Development Program'' (LDRD 12-034).
This work was supported in part by the ``Bundesministerium f\"ur
Bildung und For\-schung'', Germany.



\begin{thebibliography}{99}
%
%
\bibitem{ref:int-oam} For a recent overview and discussions of the spin
sum rule and orbital angular momentum, see the talks given at 
the INT workshop on ``Orbital Angular Momentum in QCD'', INT, Seattle, 2012,
{\tt http://www.int.washington.edu/PROGRAMS/12-49w}.
%
\bibitem{deFlorian:2008mr} 
  D.~de Florian, R.~Sassot, M.~Stratmann, and W.~Vogelsang,
  Phys.\ Rev.\ Lett.\  {\bf 101}, 072001 (2008);  
  Phys.\ Rev.\ D {\bf 80}, 034030 (2009).
%
\bibitem{Blumlein:2010rn} 
  J.~Blumlein and H.~Bottcher,
  Nucl.\ Phys.\ B {\bf 841}, 205 (2010);
  E.~Leader, A.~V.~Sidorov, and D.~B.~Stamenov,
  Phys.\ Rev.\ D {\bf 82}, 114018 (2010).
%
\bibitem{Jager:2002xm} 
  B.~Jager, A.~Schafer, M.~Stratmann, and W.~Vogelsang,
  Phys.\ Rev.\ D {\bf 67}, 054005 (2003).
%
\bibitem{Jager:2004jh} 
  B.~Jager, M.~Stratmann, and W.~Vogelsang,
  Phys.\ Rev.\ D {\bf 70}, 034010 (2004).
%
\bibitem{ref:rhic} A.~Adare {\it et al.}  [PHENIX Collaboration],
  Phys.\ Rev.\  D {\bf 76}, 051106 (2007);
  Phys.\ Rev.\ Lett.\ {\bf 103}, 012003 (2009);
  Phys.\ Rev.\  D {\bf 79}, 012003 (2009);
  B.~I.~Abelev {\it et al.}  [STAR Collaboration],
  Phys.\ Rev.\ Lett.\  {\bf 97}, 252001 (2006);
  Phys.\ Rev.\ Lett.\  {\bf 100}, 232003 (2008);
  L.~Adamczyk {\it et al.}  [STAR Collaboration],
  Phys.\ Rev.\ D {\bf 86}, 032006 (2012).
%
\bibitem{ref:starnew} P.\ Djawotho [for the STAR Collaboration], 
{\tt arXiv:1106.5769}; J.\ Phys.\ Conf.\ Ser.\ {\bf 295} (2011) 012061.
%
\bibitem{ref:dssvplus} 
  D.~de Florian, R.~Sassot, M.~Stratmann, and W.~Vogelsang,
  talk presented at ``DIS2011'', April 2011, Newport News, VA,
  {\tt arXiv:1108.3955};
  Prog.\ Part.\ Nucl.\ Phys.\  {\bf 67}, 251 (2012).
%
\bibitem{ref:rhicwp} E.C.\ Aschenauer {\it et al.}, report on ``The RHIC Spin Program:
Achievements and Future Opportunities'', BNL, 2012,
{\tt http://www.bnl.gov/npp/docs/} {\tt RHIC-Spin-WriteUp-121105.pdf}
%
\bibitem{Boer:2011fh} 
  D.~Boer {\it et al.},
  ``INT report on EIC Science'',
  {\tt arXiv:1108.1713}.
%
\bibitem{Aschenauer:2012ve} 
  E.~C.~Aschenauer, R.~Sassot, and M.~Stratmann,
  Phys.\ Rev.\ D {\bf 86} (2012) 054020.
%
\bibitem{ref:hermes}
  A.~Airapetian {\it et al.} [HERMES Collaboration],
  Phys.\ Rev.\ Lett.\  {\bf 84}, 2584 (2000);
  JHEP {\bf 1008}, 130 (2010).
%
\bibitem{ref:smc} 
  B.~Adeva {\it et al.} [Spin Muon Collaboration (SMC)],
  Phys.\ Rev.\  D {\bf 70}, 012002 (2004).
%
\bibitem{ref:compass-2had}   
  E.~S.~Ageev {\it et al.} [COMPASS Collaboration],
  Phys.\ Lett.\  B {\bf 633}, 25 (2006);
  C.~Adolph {\it et al.}  [COMPASS Collaboration],
  {\tt arXiv:1202.4064}.
%
\bibitem{ref:compass-charm}  
  M.~Alekseev {\it et al.} [COMPASS Collaboration],
  {\tt arXiv:0802.3023}; Phys.\ Lett.\  B {\bf 676}, 31 (2009).
%
\bibitem{Bojak:1998bd} 
  I.~Bojak and M.~Stratmann,
  Phys.\ Lett.\ B {\bf 433}, 411 (1998);
  Nucl.\ Phys.\ B {\bf 540}, 345 (1999)
  [Erratum-ibid.\ B {\bf 569}, 694(E) (2000)].
%
\bibitem{Merebashvili:2000ya} 
  Z.~Merebashvili, A.~P.~Contogouris, and G.~Grispos,
  Phys.\ Rev.\ D {\bf 62}, 114509 (2000)
  [Erratum-ibid.\ D {\bf 69}, 019901 (2004)];
  Phys.\ Lett.\  B {\bf 482}, 93 (2000).
%
\bibitem{Jager:2003vy} 
  B.~Jager, M.~Stratmann, and W.~Vogelsang,
  Phys.\ Rev.\ D {\bf 68}, 114018 (2003);
  Eur.\ Phys.\ J.\ C {\bf 44}, 533 (2005);
  C.~Hendlmeier, A.~Schafer, and M.~Stratmann,
  Eur.\ Phys.\ J.\ C {\bf 55}, 597 (2008).
%
\bibitem{Riedl:2009ye} 
  J.~Riedl, A.~Schafer, and M.~Stratmann,
  Phys.\ Rev.\ D {\bf 80}, 114020 (2009).
%
\bibitem{Bojak:2001fx} 
  I.~Bojak and M.~Stratmann,
  Phys.\ Rev.\ D {\bf 67}, 034010 (2003).
%
\bibitem{ref:cacciari}
  M.~Cacciari, P.~Nason and R.~Vogt,
  Phys.\ Rev.\ Lett.\  {\bf 95}, 122001 (2005).
%
\bibitem{ref:kurek}
  K.~Kurek, Habilitation Thesis,		
  National Centre of Nuclear Research, Swierk, November 2011. 
%
\bibitem{ref:franco-phd}
  C.~Franco, Ph.D. Thesis, Universidade Tecnica de Lisboa, December 2011. 
%
\bibitem{ref:franco}
  C.~Franco [for the COMPASS Collaboration], {\tt arXiv:1208.6567}.
%
%
\bibitem{Mangano:1991jk} 
  M.~L.~Mangano, P.~Nason, and G.~Ridolfi,
  Nucl.\ Phys.\ B {\bf 373}, 295 (1992).
%
\bibitem{Frixione:1993dg} 
  S.~Frixione, M.~L.~Mangano, P.~Nason, and G.~Ridolfi,
  Nucl.\ Phys.\ B {\bf 412}, 225 (1994).
%
\bibitem{ref:nlo-unpol} 
  R.~K.~Ellis and P.~Nason,
  Nucl.\ Phys.\ B {\bf 312}, 551 (1989);
  J.~Smith and W.~L.~van Neerven,
  Nucl.\ Phys.\ B {\bf 374}, 36 (1992);
  W.~Beenakker, H.~Kuijf, W.~L.~van Neerven, and J.~Smith,
  Phys.\ Rev.\ D {\bf 40}, 54 (1989);
  W.~Beenakker, W.~L.~van Neerven, R.~Meng, G.~A.~Schuler, and J.~Smith,
  Nucl.\ Phys.\ B  {\bf 351}, 507 (1991);
  P.~Nason, S.~Dawson, and R.~K.~Ellis,
  Nucl.\ Phys.\ B {\bf 327}, 49 (1989)
  [Erratum-ibid. {\bf B335}, 260 (1990)].
%
\bibitem{ref:hvbm}
  G.~'t Hooft and M.~J.~G.~Veltman,
  Nucl.\ Phys.\  B {\bf 44}, 189 (1972);
  P.~Breitenlohner and D.~Maison,
  Commun.\ Math.\ Phys.\  {\bf 52}, 11 (1977).
%
\bibitem{ref:nlo-split}
  R.~Mertig and W.~L.~van Neerven,
  Z.\ Phys.\  C {\bf 70}, 637 (1996);
  W.~Vogelsang, Phys.\ Rev.\  D {\bf 54}, 2023 (1996);
  Nucl.\ Phys.\  B {\bf 475}, 47 (1996).
%
%
\bibitem{ref:frag}
  V.~G.~Kartvelishvili, A.~K.~Likhoded, and V.~A.~Petrov,
   Phys.\ Lett.\  B {\bf 78}, 615 (1978).
%
\bibitem{ref:frag-review}
  For a review, see, e.g.,
  J.~Baines {\it et al.}, summary report of the ``Heavy Quarks Working Group''
  for the ``HERA-LHC Workshop'' proceedings, {\tt arXiv:hep-ph/0601164} and references
  therein.
%
\bibitem{ref:cteq6} 
  J.~Pumplin, D.~R.~Stump, J.~Huston, H.~L.~Lai, P.~M.~Nadolsky, and W.~K.~Tung,
  JHEP {\bf 0207}, 012 (2002).
%
\bibitem{ref:grsv} 
  M.~Gl\"uck, E.~Reya, M.~Stratmann, and W.~Vogelsang,
  Phys.\ Rev.\  D {\bf 63}, 094005 (2001).  
%
\bibitem{ref:polphoton} M.\ Gl\"uck and W.\ Vogelsang, 
  Z. Phys. C {\bf 55}, 353 (1992); C {\bf 57}, 309 (1993); 
  M.\ Gl\"uck, M.\ Stratmann, and W.\ Vogelsang, 
  Phys. Lett. B {\bf 337}, 373 (1994); 
  M.\ Stratmann and W.\ Vogelsang, 
  Phys. Lett. B {\bf 386}, 370 (1996).
%
\bibitem{ref:grvphoton} M.~Gl\"uck, E.~Reya, and A.~Vogt,
  Phys.\ Rev.\  D {\bf 46}, 1973 (1992).
%
\bibitem{ref:polww} D.\ de Florian and S.\ Frixione,
  Phys. Lett. B {\bf 457}, 236 (1999).
%
%
\bibitem{ref:compass-new} C.\ Adolph {\it et al.} [COMPASS Collaboration],
{\tt arXiv:1211.6849}.
%
\end{thebibliography}
\end{document}